\documentclass[twocolumn,prb,superscriptaddress]{revtex4}

\usepackage{graphicx}% Include figure files
\usepackage{dcolumn}% Align table columns on decimal point
\usepackage{bm}% bold math
\usepackage{amssymb, amsmath}
\usepackage{color}
\usepackage{slashed} 

\begin{document}
\title{Superconductivity in systems exhibiting the Altshuler-Aronov anomaly}

\author{Branislav Rabatin}

\altaffiliation[Present address: ]{Department of Physics, Florida
  State University, Tallahassee, Florida 32306, USA}

\affiliation{Department of Experimental Physics, Comenius University,
  Mlynsk\'{a} Dolina F2, 842 48 Bratislava, Slovakia}

\affiliation{Institute of Informatics, Slovak Academy of Sciences,
  D\'{u}bravsk\'{a} cesta 9, 845 07 Bratislava, Slovakia}

\author{Richard Hlubina}
\affiliation{Department of Experimental Physics, Comenius University,
  Mlynsk\'{a} Dolina F2, 842 48 Bratislava, Slovakia}

\begin{abstract}
Making use of generalized Eliashberg equations, we describe the
Altshuler-Aronov (AA) effect and superconductivity on equal footing.
We derive explicit expressions for the Coulomb pseudopotential in 3D,
taking into account also the anomalous diffusion.  We present a full
numerical solution for two normal-state and two anomalous
self-energies.  In the normal state, we amend the known results for
the purely electronic AA effect; with electron-phonon coupling turned
on, we find additional anomalies in the density of states close to the
phonon energy.  We study how the critical temperature and density of
states of strongly disordered 3D superconductors change with
normal-state resistivity. We find that the type of transition from the
superconducting to the insulating state depends on the strength of
electron-phonon coupling: at weak coupling there exists an
intermediate normal state, whereas at strong coupling the transition
is direct.
\end{abstract}
\maketitle

\section{Introduction}
One of the central questions which have not been answered yet in the
context of high-temperature superconductivity is that about the origin
of the so-called pseudogap. Under the pseudogap, a complex of
phenomena in the non-superconducting state of lightly doped cuprates
is understood, among which a prominent place is taken by the
suppression of the number of states in the vicinity of the Fermi
level, documented by thermodynamic as well as spectroscopic methods
\cite{Keimer15}.

Several candidate explanations have been proposed for the pseudogap in
the cuprates, which can be classified into two large groups. In the
first type of theories, the pseudogap is understood as a consequence
of some symmetry-breaking phase transition leading to the formation of
a ``competing order''. In moderate theories of this type it is assumed
that the competing order is not static, but only fluctuating. The
second type of theories views the pseudogap as a consequence of the
proximity of the cuprates to a Mott insulating state. It is the latter
type of theories which motivates the present work.

If one assumes that the superconductor is close to being insulating,
then there are again two pictures of how the pseudogap may arise,
which have been widely studied also in systems other (and presumably
simpler) than the cuprates \cite{Gantmakher10,Larkin99}. The first
(so-called bosonic) picture builds on the observation that the
superfluid stiffness should be small in the vicinity of the insulating
state, leading to strong phase fluctuations. Within this picture, the
pseudogap appears due to the presence of Cooper pairs which have not
condensed into a single macroscopically occupied state
\cite{Doniach81}. The second (so-called fermionic) picture starts from
the observation that the screening of the Coulomb interactions should
become progressively weaker and weaker as the insulating state is
approached, and therefore the consequences of such interactions should
become visible in the metallic state, irrespective of whether it
ultimately becomes superconducting at low temperatures
\cite{Finkelstein94}.

On the experimental side, pseudogap has been observed in several
low-temperature superconductors which are close to being insulating.
Among the first observations predating the cuprate era were those in
granular aluminum \cite{Dynes81} and in the alloy Nb$_{1-x}$Si$_x$
\cite{Hertel83}. More recently pseudogap has been observed in very
diverse systems such as TiN \cite{Sacepe10}, InO$_x$ \cite{Sacepe11},
NbN \cite{Mondal11}, BaPb$_{1-x}$Bi$_x$O$_3$ \cite{Luna14}, and
Cu$_x$TiSe$_2$ \cite{Luna15}.

A mere observation of the pseudogap does not allow us to distinguish
between the bosonic and fermionic scenaria, and therefore more
quantitative predictions of the theory are needed. Far away from the
transition, a perturbative calculation within the fermionic scenario
due to Altshuler and Aronov (AA) \cite{Altshuler79} suggests that the
density of states $N(\omega)$ in the vicinity of the Fermi level of a
3D metal should be suppressed according to
\begin{equation}
N(\omega)=
N(0)\left[1+\sqrt{\omega/\Delta_{\rm AA}}\right].
\label{eq:dos_aa}
\end{equation}
In several papers the formula Eq.~\eqref{eq:dos_aa} has in fact been
observed experimentally \cite{Dynes81,Hertel83,Luna14,Luna15}.
However, as we will show, the observed magnitude of $\Delta_{\rm AA}$
is orders of magnitude smaller than predicted by a straightforward
extrapolation of the AA theory.

In this paper we will demonstrate that a generalization of the AA
theory due to Anderson, Muttalib, and Ramakrishnan (AMR)
\cite{Anderson83}, which takes into account the scale-dependence of
the diffusion constant predicted by the scaling theory of localization
\cite{Abrahams79}, also leads to a density of states of the fom
Eq.~\eqref{eq:dos_aa}, but with an energy scale $\Delta_{\rm AA}$
which is compatible with the experiments. A similar observation has
been made also previously \cite{Lee82,Muttalib83}.

Our next goal is to apply the AA theory, as modified by AMR, to the
superconducting state and to check whether the results of
\cite{Dynes81,Hertel83,Luna14,Luna15} can in fact be explained within
the fermionic theory. To this end we will study the generalized
Eliashberg equations in the exact eigenstate basis. Our approach is
quite similar in spirit to the pioneering work of Belitz
\cite{Belitz87}, the main technical difference being that we work on
the imaginary axis. Moreover, unlike Belitz, we will present the
results of a full numerical solution of the Eliashberg equations. Thus
we have access not only to thermodynamics, but, after analytic
continuation, also to the superconducting density of states.

The outline of this paper is as follows. Following AMR, in Section~2
we determine the energy dependence of the Coulomb pseudopotential
$\mu(\varepsilon)$. We show that in addition to the logarithmic regime
at intermediate $\varepsilon$ discovered by AMR in strongly disordered
systems, at the lowest $\varepsilon$ the pseudopotential always varies
as $\sqrt{\varepsilon}$, but the relevant energy scale varies by
orders of magnitude between the weakly and strongly disordered
systems. In Section~3 we write down the Eliashberg equations in the
exact eigenstate basis and in Section~4 we show that their solution in
the normal state always leads to a density of states of the form
Eq.~\eqref{eq:dos_aa} in the low-energy limit. Furthermore we show
that, in the strongly disordered limit, the energy scale $\Delta_{\rm
  AA}$ can become arbitrarily small, in qualitative agreement with
\cite{Dynes81,Hertel83}. We also show here that, in presence of the AA
effect, coupling to phonons leads to additional features of the
density of states (in the nonsuperconducting state) close to the
phonon energy.  Finally, in Section~5 we present the results of the
numerical solution of the Eliashberg equations in the superconducting
state.

\section{Coulomb pseudopotential}
Within the Eliashberg theory, the exchange contribution to the bare
Coulomb pseudopotential is calculated as a Fermi-surface average of
the screened Coulomb interaction. If we consider the static screening
with inverse screening length $k_s$, one can show readily that the
formula
\begin{equation}
\mu(\varepsilon)=\frac{e^2}{2\pi^3\epsilon_0}
\int_0^{2k_F}\frac{dq q^2}{q^2+k_s^2}
\frac{\hbar D_q q^2}{(\hbar D_q q^2)^2+\varepsilon^2},
\label{eq:coulomb_def}
\end{equation}
wherein we take $D_q=2v_F/(\pi q)$ with $v_F$ the Fermi velocity, does
lead - at zero energy transfer $\varepsilon=0$ - to the well-known
Coulomb pseudopotential $\mu_0=\frac{\alpha}{2\pi}\ln(1+\pi/\alpha)$
of a clean system with isotropic quadratic dispersion \cite{Morel62}.
Here we have introduced the ``fine structure'' constant
$\alpha=e^2/(4\pi\epsilon_0\hbar v_F)$ of the electron gas
\cite{note_rs} and we have made use of the relation
$2k_F/k_s=(\pi/\alpha)^{1/2}$. Note that in a typical metal
$\alpha\sim 1$.

On the other hand, as shown by \cite{Altshuler79,Belitz87},
Eq.~\eqref{eq:coulomb_def} is applicable also to weakly disordered
systems, and in this case one has to take $D_q=D_0$, where
$D_0=\tfrac{1}{3}v_F\ell$ is the diffusion constant of the dirty
system characterized by the mean free path $\ell$. This result can be
most simply shown in the basis of exact Hartree-Fock eigenstates of
the disordered system, in which the exchange contribution to the
self-energy $\chi(\varepsilon)$ of an eigenstate with bare energy
$\varepsilon$ reads as $\chi(\varepsilon)=-\int d\varepsilon^\prime
\mu(\varepsilon-\varepsilon^\prime) f(\varepsilon^\prime)$, where
$f(x)$ is the Fermi function. We deliberately neglect all Hartree
contributions to the self-energy, although in a more complete
treatment of a disordered system they may be present \cite{Belitz87}.

The goal of this Section, which represents a generalization of the
insightful AMR paper \cite{Anderson83}, is to study the evolution of
the Coulomb pseudopotential $\mu(\varepsilon)$ with the amount of
disorder: from the clean case, via weakly disordered systems, up to
the strongly disordered (but still metallic) case.

Before proceeding, it is useful to introduce a sharp criterion which
enables us to distinguish between weak and strong disorder. It is well
known that the naive formula for the conductivity, $\sigma=g_0
k_F^2\ell$ with $g_0=e^2/(3\pi^2\hbar)$ and what AMR call the local
mean free path $\ell$, should not be applied too close to the
localized regime, because in that case localization corrections enter
the expression for $\sigma(\ell)$. Instead, the scaling theory of
localization \cite{Abrahams79} suggests to introduce a different length
scale $L_s$ such that $\sigma=g_c/L_s$, where $g_c\sim g_0$ is the
critical conductance. Following AMR, we will call systems with
$L_s<\ell$ weakly disordered, and those with $L_s>\ell$ strongly
disordered. It turns out that it is advantageous to discuss these two
cases separately.

\subsection{Weak disorder, $L_s<\ell$}
In the weakly disordered case one can identify two qualitatively
different contributions to Eq.~\eqref{eq:coulomb_def}. Namely, at
short distances (i.e. for wave-vectors $q^\ast<q<2k_F$), the electron
motion should be ballistic and we should therefore use the clean-limit
expression \cite{note_ballistic} $D_q=2v_F/(\pi q)$. On the other hand,
at long length scales (i.e. for wave-vectors $0<q<q^\ast$), the
electron motion is diffusive and we should take $D_q=D_0$.  Note that
in doing so, we reduce the diffusion constant with respect to its
ballistic value, which in turn leads to an increase of the Coulomb
pseudopotential. In the spirit of AMR, we assume that the short- and
long-distance forms are valid up to the cross-over scale
$q^\ast$. Requiring that the function $D_q$ is continuous leads us
then to the identification $q^\ast=6/(\pi \ell)$. Note that the
inequality $q^\ast<2k_F$ implies that we have to require
$k_F\ell>3/\pi$.

With this choice of the function $D_q$, the integral in
Eq.~\eqref{eq:coulomb_def} can be taken exactly, but it leads to a
bulky formula. We find that the result can be written with good
accuracy by the following expression:
\begin{equation}
\mu(\varepsilon)=
\left\{
\begin{array}{ll}
  \mu_0+\frac{1}{2(k_F\ell)^2}
  \left[1-\delta-\sqrt{\frac{\varepsilon}{\Gamma}}\right],
&
\varepsilon<\varepsilon_{\rm max},
\\
\mu_0, & \varepsilon>\varepsilon_{\rm max},
\end{array}
\right.
\label{eq:coulomb_weak}
\end{equation}
where we have introduced the energy $\Gamma=2\varepsilon_F/(3k_F\ell)$
and the dimensionless number $\delta$
$$
\delta=\frac{\alpha(k_F\ell)^2}{6}
\ln\left[1+\frac{1}{\alpha(k_F\ell)^2}\right].
$$
Note that $\delta<0.17$, i.e. $\delta$ is always small.  The energy
scale $\varepsilon_{\rm max}$ can be found by requiring that
$\mu(\varepsilon)$ should not drop below its value $\mu_0$ in the
clean metal.

Note that Eq.~\eqref{eq:coulomb_weak} looks reasonable: the Coulomb
pseudopotential $\mu(\varepsilon)$ of a disordered system is larger
than $\mu_0$, at small energy transfers it exhibits the expected
$\sqrt{\varepsilon}$-dependence, and in the clean limit
$k_F\ell\rightarrow\infty$ it reduces to $\mu_0$. In Section~4 we will
demonstrate that the well-known AA depression of the density of states
at the Fermi level \cite{Altshuler79} is a simple consequence of
Eq.~\eqref{eq:coulomb_weak}.

\subsection{Strong disorder, $\ell< L_s$}
Also in this case we will construct, following AMR, the simplest
scale-dependent diffusion coefficient $D_q$ which is consistent with
the known physical constraints.  Let us start at the largest
length-scales, where, as noted by AMR, the macroscopic diffusion
constant of a strongly disordered system with $\ell< L_s$ is reduced
from its local estimate $D_0$ to $D_0\ell/L_s$. Therefore in the
region $0<q<L_s^{-1}$ we will assume that $D_q=D_0\ell/L_s$.  At
intermediate length scales AMR have identified a region of anomalous
diffusion \cite{note_diffusion}, where $D_q=D_0 q\ell$ and the
diffusion constant increases with decreasing length scale, ultimately
approaching its local limit $D_0$. This functional form will therefore
be assumed to be valid at momenta $L_s^{-1}<q<\ell^{-1}$.  Since the
diffusion constant of a dirty system can not exceed its local limit,
at still shorter length scales $\ell^{-1}<q<q^\ast$ we have to assume
that $D_q=D_0$, until ultimately at the shortest length scales
$q^\ast<q<2k_F$ the electrons move ballistically and therefore
$D_q=2v_F/(\pi q)$.

With the above choice of the function $D_q$, the integral in
Eq.~\eqref{eq:coulomb_def} can again be taken exactly. The result can
be reasonably well described by the following function:
\begin{equation}
\mu(\varepsilon)=  
\begin{cases}
\mu_0+\frac{1}{2(k_F\ell)^2}
\left[1+\ln\left(\frac{L_s}{\ell}\right)-\delta
  -\sqrt{\frac{\varepsilon}{\varepsilon^\ast}}\right],\;
\varepsilon<\varepsilon^\ast,
\\
\mu_0+\frac{1}{2(k_F\ell)^2}
\left[\frac{1}{3}\ln\left(\frac{\Gamma}{\varepsilon}\right)-\delta\right],
\hspace{0.55cm}
\varepsilon^\ast<\varepsilon<\varepsilon_{\rm max},
\\
\mu_0,\hspace{5.05cm}
\varepsilon_{\rm max}<\varepsilon,
\end{cases}  
\label{eq:coulomb_strong}
\end{equation}
where $\varepsilon^\ast=(\ell/L_s)^3\times\Gamma$ is a new energy
scale. Note that in a strongly disordered metal
$\varepsilon^\ast\ll\Gamma$. Requiring that $\mu(\varepsilon)$ is
continuous we find $\varepsilon_{\rm max}=\Gamma e^{-3\delta}$.

When Eq.~\eqref{eq:coulomb_strong} is compared with the result
Eq.~\eqref{eq:coulomb_weak} for the weakly disordered case, one can
notice that the low-energy enhancement of the Coulomb pseudopotential
is much larger in the present case. There are two reasons for this:
first, the factor $k_F\ell\sim 1$ is much smaller than $k_F\ell\gg 1$
in the weakly disordered case. Second and less trivially, due to
anomalous diffusion, at intermediate energy transfers
$\varepsilon^\ast<\varepsilon<\Gamma$ the Coulomb pseudopotential
exhibits a large logarithmic increase, in qualitative agreement with
the result of AMR.

It should be pointed out that at the lowest energy transfers
$\varepsilon<\varepsilon^\ast$, which have not been considered by AMR,
the Coulomb pseudopotential $\mu(\varepsilon)$ still exhibits the
standard AA-type behavior, but the associated energy scale is
$\varepsilon^\ast$ instead of $\Gamma$, i.e. it may be much smaller
than in the weakly disordered systems. This has observable
consequences, as explained in Section~4.

\section{Eliashberg equations}
In the basis of exact eigenstates of the disordered system, the
Eliashberg equations can be written in the imaginary-time Nambu-Gorkov
formalism in a very compact form
$$
\hat{\Sigma}(\varepsilon,\omega)=
T\sum_{\omega^\prime}\int d\varepsilon^\prime
[-\mu(\varepsilon^\prime-\varepsilon) +g(\omega^\prime-\omega)]
\tau_3 \hat{G}(\varepsilon^\prime,\omega^\prime) \tau_3,
$$
where $\tau_3$ is the Pauli matrix and the $2\times 2$ matrices
$\hat{\Sigma}(\varepsilon,\omega)$ and $\hat{G}(\varepsilon,\omega)$
are the self-energy and the Green function for a time-reversal related
pair of eigenstates characterized by bare energy $\varepsilon$;
$\omega$ is the Matsubara frequency.  In what follows we do
distinguish between energy and frequency; however, both will be
measured in the same units, i.e. we set $\hbar=1$.  Note that in a
disordered system $\varepsilon$ plays the same role as momentum ${\bf
  k}$ in a clean system. That is also the reason why the Coulomb
pseudopotential (in a theory with static screening) is a function of
transferred energy and not frequency.

The Eliashberg equations describe the contributions of self-consistent
rainbow diagrams to the self-energy, where the interaction lines are
either due to screened Coulomb interactions described by the Coulomb
pseudopotential $\mu(\varepsilon)$ introduced in the previous Section,
or due to electron-electron interactions generated by the exchange of
phonons and described by the function $g(\omega)$. In what follows, we
will assume a simple Debye model for the phonons, and the resulting
function $g(\omega)$ reads as \cite{Bzdusek15}
\begin{equation}
g(\omega)=\lambda
\left[1-\frac{\omega^2}{\Omega^2}
  \ln\left(1+\frac{\Omega^2}{\omega^2}\right)\right],
\label{eq:phonons}
\end{equation}
where $\lambda$ is the dimensionless electron-phonon coupling and
$\Omega$ is the Debye energy. Following the arguments of AMR
\cite{Anderson83,Keck76}, in what follows we neglect the effect of
disorder on $\lambda$, since we intend to concentrate on strongly
disordered superconductors, where the effect of the Coulomb
pseudopotential should dominate. For the same reason we keep
neglecting all possible Hartree-type contributions to the self-energy.

The most general ansatz for the self-energy
$\hat{\Sigma}(\varepsilon,\omega)$ can be written as
\begin{equation}
\hat{\Sigma}(\varepsilon,\omega)
=\Sigma(\varepsilon,\omega)\tau_0
+\chi(\varepsilon,\omega)\tau_3
+\Phi(\varepsilon,\omega)\tau_1,
\label{eq:self_ansatz}
\end{equation}
where $\tau_i$ are the Pauli matrices and the functions
$\Sigma(\varepsilon,\omega)$ and $\Phi(\varepsilon,\omega)$ are the
normal and anomalous self-energies, respectively. In clean
particle-hole symmetric metals, the $\tau_3$-component of the
self-energy can be ignored, because the Coulomb pseudopotential can be
taken as energy-independent. However, as explained in the previous
Section, in disordered systems $\mu(\varepsilon)$ is not a constant
function, and therefore in addition to $\Sigma(\varepsilon,\omega)$
and $\Phi(\varepsilon,\omega)$, also the function
$\chi(\varepsilon,\omega)$ has to be determined self-consistently.
Moreover, the $\varepsilon$-dependence can not be simply ignored as in
the clean case.  These points have been emphasized by Belitz early
on \cite{Belitz87}.

In what follows it is useful to define also the functions
$Z(\varepsilon,\omega)=1+\Sigma(\varepsilon,\omega)/(i\omega)$ and
$R(\varepsilon,\omega)=1+\chi(\varepsilon,\omega)/\varepsilon$.
Inserting the ansatz Eq.~\eqref{eq:self_ansatz} into the Eliashberg
equations and making use of the fact that the functions
$\mu(\varepsilon)$ and $g(\omega)$ are even, one can show that also
$Z$, $R$ and $\Phi$ can be chosen as even functions of
$\varepsilon,\omega$.  Making use of this observation one finds
readily that $\Sigma(\omega)$ and $Z(\omega)$ are independent of
$\varepsilon$, and similarly $\chi(\varepsilon)$ and $R(\varepsilon)$
do not depend on $\omega$.  The Eliashberg equations for a system at
temperature $T$ therefore simplify to the following form:
\begin{eqnarray}
\Sigma(\omega)&=&T\sum_{\omega^\prime}\int 
\frac{d\varepsilon^\prime
g(\omega^\prime-\omega)i\omega^\prime Z^\prime}
{(\omega^\prime Z^\prime)^2+(\varepsilon^\prime R^\prime)^2
  +\Phi^2(\varepsilon^\prime,\omega^\prime)},
\label{eq:sigma}
\\
\chi(\varepsilon)&=&T\sum_{\omega^\prime}\int
\frac{d\varepsilon^\prime
\mu(\varepsilon^\prime-\varepsilon)\varepsilon^\prime R^\prime}
{(\omega^\prime Z^\prime)^2+(\varepsilon^\prime R^\prime)^2
  +\Phi^2(\varepsilon^\prime,\omega^\prime)},
\label{eq:chi}
\\
\phi(\omega)&=&T\sum_{\omega^\prime}\int
\frac{d\varepsilon^\prime
g(\omega^\prime-\omega)\Phi(\varepsilon^\prime,\omega^\prime)}
{(\omega^\prime Z^\prime)^2+(\varepsilon^\prime R^\prime)^2
  +\Phi^2(\varepsilon^\prime,\omega^\prime)},
\label{eq:phi}
\\
\psi(\varepsilon)&=&T\sum_{\omega^\prime}\int 
\frac{d\varepsilon^\prime
\mu(\varepsilon^\prime-\varepsilon)\Phi(\varepsilon^\prime,\omega^\prime)}
{(\omega^\prime Z^\prime)^2+(\varepsilon^\prime R^\prime)^2
  +\Phi^2(\varepsilon^\prime,\omega^\prime)},
\label{eq:psi}
\end{eqnarray}
where we have introduced the abbreviations $Z^\prime=Z(\omega^\prime)$
and $R^\prime=R(\varepsilon^\prime)$.  We have also observed that the
anomalous self-energy can be written as
$\Phi(\varepsilon,\omega)=\phi(\omega)-\psi(\varepsilon)$.

In the rest of this paper we will be concerned with the solution of
Eqs.~(\ref{eq:sigma},\ref{eq:chi},\ref{eq:phi},\ref{eq:psi}) with
interactions given by
Eqs.~(\ref{eq:coulomb_weak},\ref{eq:coulomb_strong},\ref{eq:phonons}).
Both in the Matsubara frequency space, as well as in the energy space
we will assume that there is a finite cut-off $\Lambda$ which
restricts the studied states to the vicinity of the Fermi energy,
$|\omega|,|\varepsilon|\leq \Lambda$. We will take $\Lambda$ much
larger than the Debye energy $\Omega$, in order to have a valid
description of the electron-phonon interaction.

In the special case when $\mu(\varepsilon)$ is a constant one can
easily observe that Eq.~\eqref{eq:chi} implies that $\chi=0$ and one
ends up with the usual Eliashberg equations. The Coulomb
pseudopotential enters only Eq.~\eqref{eq:psi} in this case. Strictly
speaking, we should not assume that it equals the bare value $\mu_0$,
since our cutoff $\Lambda$ is much smaller than the Fermi energy (or
bandwidth), and we should rather use an appropriately renormalized
value. Nevertheless, since this is a minor correction, we have decided
to use the bare value of $\mu_0$ instead. On the other hand, we
emphasize that the renormalization of the Coulomb interaction from the
scale $\Lambda$ to the phonon scale $\Omega$ is implicitly present in
our self-consistent calculations.

Once the Eliashberg equations are solved, the Matsubara Green function
$\hat{G}(\varepsilon,\omega)$ of the superconductor can be determined
from the Dyson equation
$$
\hat{G}^{-1}(\varepsilon,\omega)=i\omega\tau_0-\varepsilon\tau_3
-\hat{\Sigma}(\varepsilon,\omega).
$$
The density of states $N(\omega)$ in the superconducting state
can be obtained from the textbook formula
\begin{equation}
  N(\omega)=-\frac{1}{\pi} N_0\int d\varepsilon
  {\rm Im} G_{11}^{\rm R}(\varepsilon,\omega),
  \label{eq:dos_def}
\end{equation}
where $G_{11}^{\rm R}(\varepsilon,\omega)$ is the upper left component
of the retarded Green function $\hat{G}^{\rm R}(\varepsilon,\omega)$,
and $N_0$ is the density of the bare levels $\varepsilon$.

\section{Normal state}
In this Section we will investigate the implications of the Eliashberg
equations for the normal-state properties of disordered metals. Since
in the normal state $\Phi=0$, we have to solve
Eqs.~(\ref{eq:sigma},\ref{eq:chi}) for the self-energies
$\chi(\varepsilon)$ and $\Sigma(\omega)$.  We will be especially
interested in the tunneling density of states. We will start by
discussing the case when the electron-phonon coupling is turned off
and we will show that the AA anomaly exhibits novel features in the
limit of strong disorder. Next we will show how switching on a finite
electron-phonon interaction leads to additional structure in the
density of states.

\subsection{Systems without electron-phonon coupling}
In this case the self-energy due to phonons vanishes,
$\Sigma(\omega)=0$ and $Z=1$. Assuming a sufficiently large cut-off
$\Lambda$, the sum over the Matsubara frequencies in
Eq.~\eqref{eq:chi} can be performed explicitly and we find
a self-consistent equation for the self-energy $\chi(\varepsilon)$,
$$
\chi(\varepsilon)=\frac{1}{2}\int_{-\Lambda}^{\Lambda}
d\varepsilon^\prime \mu(\varepsilon^\prime-\varepsilon)
\tanh\frac{\varepsilon^\prime+\chi(\varepsilon^\prime)}{2T}.
$$
In order to proceed, let us take into account that the Coulomb
pseudopotential can be written as
$\mu(\varepsilon)=\mu_0+\delta\mu(\varepsilon)$, where the function
$\delta\mu(\varepsilon)$ vanishes for $\varepsilon>\varepsilon_{\rm
  max}$. For the sake of simplicity let us specialize to the case
of $T=0$. A simple calculation shows that in this case
$$
\chi(\varepsilon)=\int_0^\varepsilon dE \delta\mu(E),
$$
a result which is valid for $\varepsilon<\varepsilon_{\rm max}$. On the
other hand, for $\varepsilon>\varepsilon_{\rm max}$ we find that
$\chi(\varepsilon)=\chi(\varepsilon_{\rm max})$ is a constant.

From Eq.~\eqref{eq:dos_def} it follows that the density of states
$N(\omega)$ of an interacting disordered system is given by
$$
N(\omega)=N_0 \int dE \delta\left[E+\chi(E)-\omega\right]
=\frac{N_0}{1+\delta\mu(E_0)},
$$
where $E_0$ is the solution of the equation $E_0+\chi(E_0)=\omega$.

In the weakly disordered regime where Eq.~\eqref{eq:coulomb_weak}
applies we thus find that the density of states in the low-frequency
limit $\omega\ll\Gamma$ can be described (to order
$\sqrt{\omega/\Gamma}$) by Eq.~\eqref{eq:dos_aa} with
$$
N(0)=\frac{N_0}{1+\frac{1-\delta}{2(k_F\ell)^2}},
\quad
\Delta_{\rm AA}=\frac{8}{3}
\left[k_F\ell+\frac{1-\delta}{2k_F\ell}\right]^3 \varepsilon_F,
$$
a well-known result due to Altshuler and Aronov \cite{Altshuler79}.
However, since throughout the weakly disordered regime we have
$1\lesssim k_F\ell$, the AA energy scale $\Delta_{\rm AA}$ is at least
of order $\varepsilon_F$, and therefore not directly observable on the
meV scale of typical tunneling experiments. This suggests that the
experimental findings of Refs.~\cite{Dynes81,Hertel83,Luna14,Luna15}
can not be explained by a straightforward application of
Altshuler-Aronov physics.

On the other hand, in the strongly disordered regime, the energy scale
$\Delta_{\rm AA}$ can be reduced substantially. In fact, for
$\omega\ll\varepsilon^\ast$ the density of states can be again written
in the form of Eq.~\eqref{eq:dos_aa}, and from
Eq.~\eqref{eq:coulomb_weak} it follows that
\begin{eqnarray*}
N(0)&=&\frac{N_0}{1+\frac{1-\delta+\ln(L_s/\ell)}{2(k_F\ell)^2}},
\\
\Delta_{\rm AA}&=&\frac{8}{3}
\left[k_F\ell+\frac{1-\delta+\ln(L_s/\ell)}{2k_F\ell}\right]^3
\left(\frac{\ell}{L_s}\right)^3\varepsilon_F.
\end{eqnarray*}
Note that in the strongly disordered regime $k_F\ell\sim 1$. Since we
can write $L_s/\ell=\rho/\rho_c$ where $\rho_c=\ell/g_c$, in the
immediate vicinity of the metal-insulator transition (where the
resistivity $\rho$ blows up) the AA energy scale $\Delta_{\rm AA}$ can
become arbitrarily small, $\Delta_{\rm AA}\propto
[\ln(\rho/\rho_c)/(\rho/\rho_c)]^3$, and this scaling is not
inconsistent with the scaling found experimentally in
Refs.~\onlinecite{Dynes81,Hertel83}. A similar result for the energy
scale $\Delta_{\rm AA}$, except for the logarithmic correction, has
been found previously \cite{Lee82,Muttalib83}.

As regards the density of states right at the Fermi energy, $N(0)$, in
perturbative calculations it is typically identified with the bare
value $N_0$ \cite{Muttalib83}. Also in our self-consistent calculation
$N(0)$ differs only weakly from the bare value $N_0$, if the system is
weakly disordered. However, in the strongly disordered regime we find
that $N(0)$ becomes heavily suppressed when $\rho\rightarrow \infty$
and the insulating state is approached, and it varies ultimately as
$N(0)\sim N_0/\ln(\rho/\rho_c)$. It is worth pointing out that the
ratio $N(0)/N_0$ is measurable and experiments with 2D systems
\cite{Valles89} do find that $N(0)/N_0<1$.

\begin{figure}[t]
  \includegraphics[width = 7 cm]{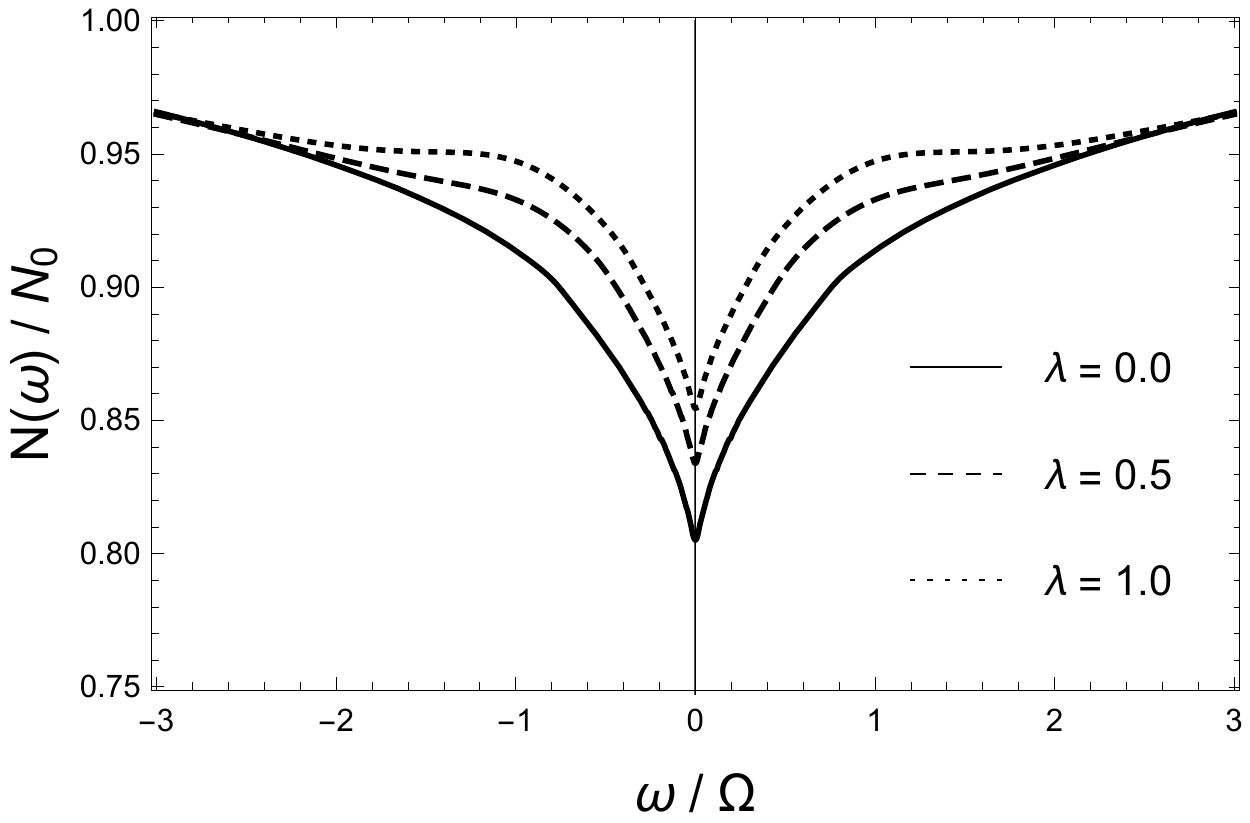}
  \caption{Density of states of a strongly disordered AA metal with
    $k_F\ell=1.8$ and $L_s/\ell=3$ in the normal state at temperature
    $T=0.01\Omega$ for three values of the electron-phonon coupling
    constant $\lambda$.}
\label{fig:dos_n}
\end{figure}

Finally, for the sake of completeness let us note that in the limit
$L_s/\ell\rightarrow\infty$ the density of states exhibits a
logarithmic correction in the limit of small frequencies:
$$
N(\omega)=
\frac{N_0}{1+\frac{1}{6(k_F\ell)^2}\ln\frac{\Gamma}{6(k_F\ell)^2\omega}}.
$$
Logarithmic scaling of the density of states in the critical regime
has been found also earlier \cite{Lee82,Muttalib83}.

\subsection{Finite electron-phonon coupling}
For a finite coupling between the electrons and phonons, an analytic
solution can be found for a constant Coulomb pseudopotential, since in
this case $\chi=0$ and $R=1$. If we furthermore assume that $T=0$ and
$\omega\ll \Lambda$, a standard calculation shows that the real part
of the retarded wave-function renormalization is
$$
{\rm Re}Z(\omega)-1=\frac{\lambda}{3}
\left[1+
\frac{\Omega}{\omega}\ln\left|\frac{\omega+\Omega}{\omega-\Omega}\right|
+\frac{\omega^2}{\Omega^2}\ln\left|1-\frac{\Omega^2}{\omega^2}\right|
\right],  
$$
which, inter alia, implies the well-known result for the mass enhancement
$Z(0)=1+\lambda$.

\begin{figure}[t]
  \includegraphics[width = 7 cm]{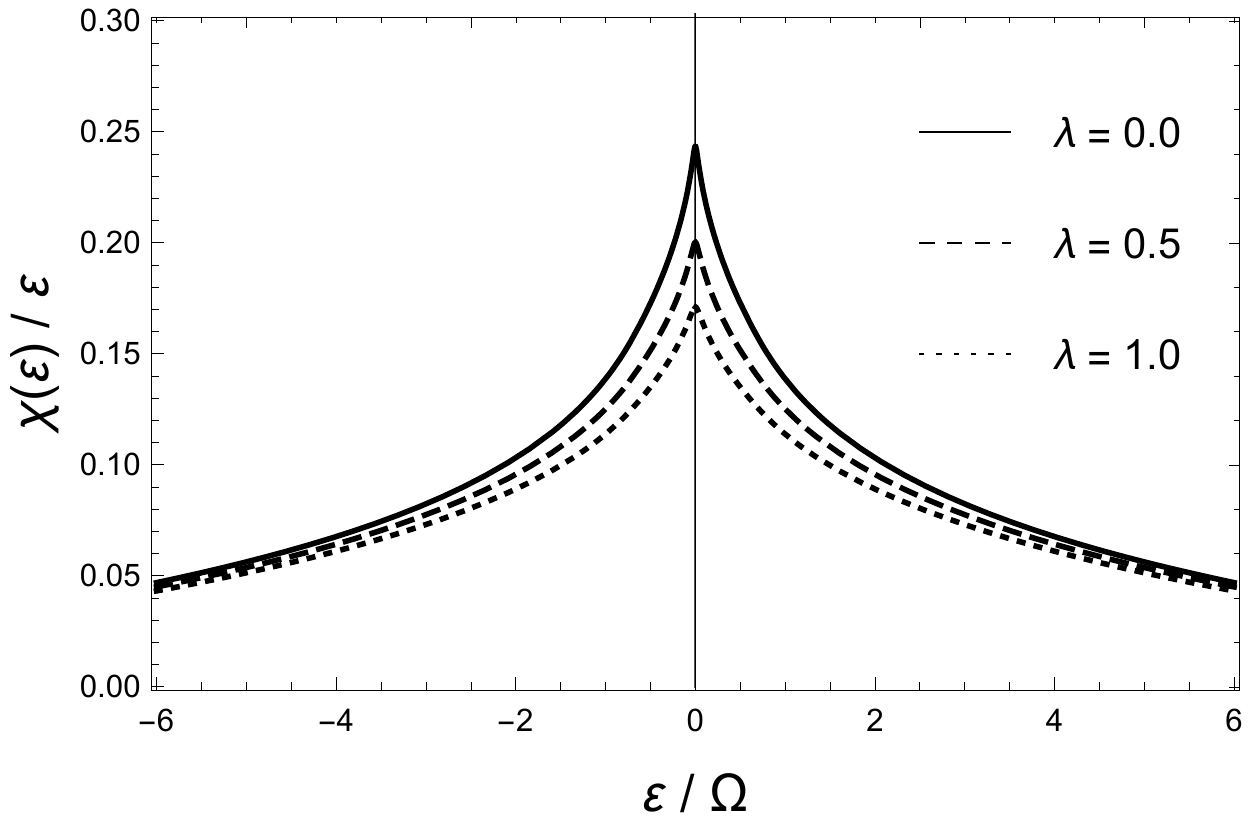}
  \caption{Normalized self-energy $\chi(\varepsilon)/\varepsilon$ for
    the same set of parameters as used in Fig.~\ref{fig:dos_n}.}
\label{fig:anticorrelation}
\end{figure}

In presence of both, a finite electron-phonon coupling $\lambda$ and
an energy-dependent Coulomb pseudopotential $\mu(\varepsilon)$, we
have solved the coupled Eqs.~(\ref{eq:sigma},\ref{eq:chi})
numerically. The analytic continuation from the Matsubara frequencies
to the real axis has been carried out by means of the Pad\'{e}
approximation \cite{Beach00}.

As a typical example of the results which we find, in
Fig.~\ref{fig:dos_n} we present the density of states calculated using
Eq.~\eqref{eq:dos_def} for a strongly disordered metal with
$k_F\ell=1.8$ (close to the critical value) and $L_s/\ell=3$.  For the
fine-structure constant we take $\alpha=1.3$, implying that the bare
Coulomb pseudopotential of the clean system is $\mu_0\approx 0.25$.
For this choice of parameters we find that the dimensionless number
$\delta\approx 0.15$.

Throughout this paper, energies will be measured in units of the Debye
energy $\Omega$. For the Fermi energy and the cutoff we take
$\varepsilon_F=50\Omega$ and $\Lambda=10\Omega$, respectively, so that
the set of inequalities $\Omega\ll\Lambda\ll\varepsilon_F$ is well
satisfied.  For our choice of parameters we have $\Gamma=
2\varepsilon_F/(3 k_F\ell)\approx 18.5\Omega$ and $\varepsilon_{\rm
  max}=\Gamma e^{-3\delta}\approx 11.8\Omega$. This implies that
essentially the whole anomalous part of the Coulomb pseudopotential is
within the cutoff, except for a small tail which can be neglected.

Figure~\ref{fig:dos_n} shows that, without coupling to phonons
(i.e. for $\lambda=0$), the density of states exhibits a strong
AA-type singularity at low frequencies, as well as a feature close to
the energy scale $\varepsilon^\ast\approx 0.69\Omega$, as should have
been expected. When a finite $\lambda$ is turned on, two new effects
become apparent.

\begin{figure}[t!]
\includegraphics[width = 7 cm]{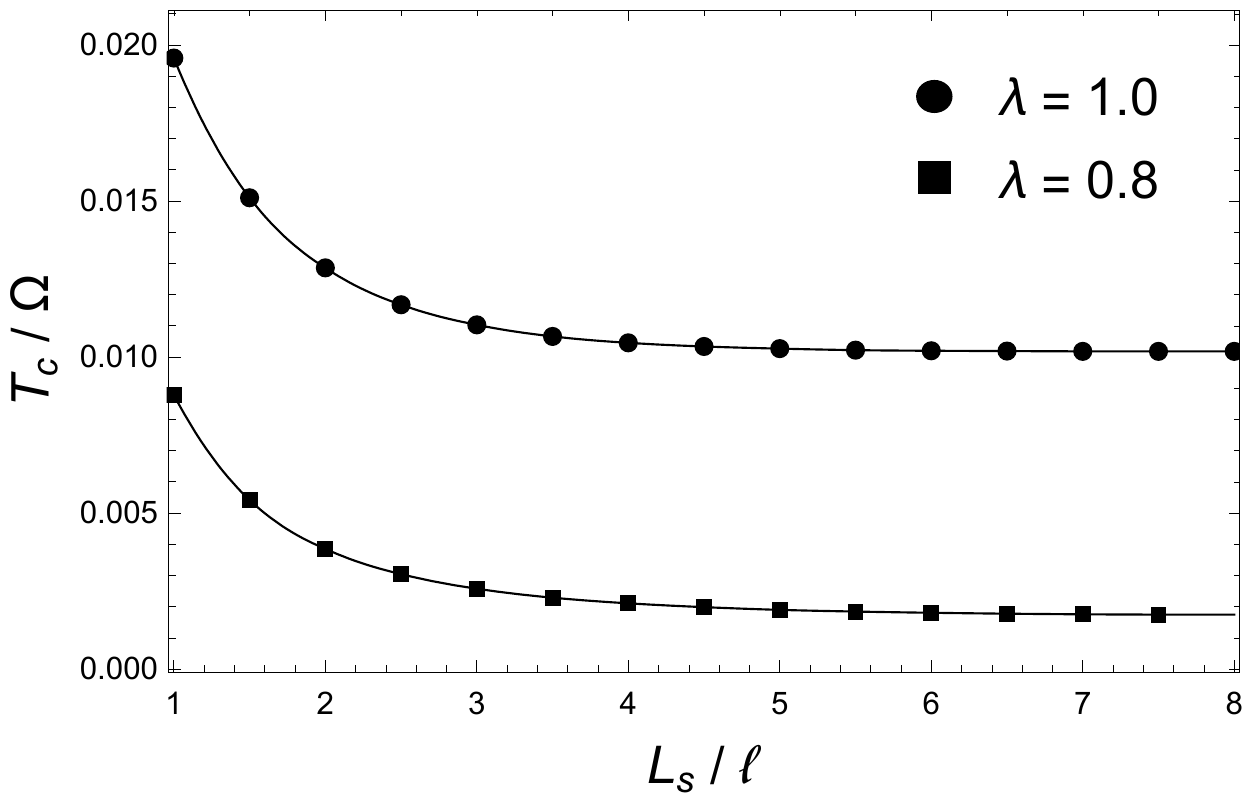}
\caption{Critical temperature $T_c$ of strongly disordered AA
  superconductors with two different electron-phonon coupling
  constants $\lambda$ as a function of $L_s/\ell=\rho/\rho_c$, where
  $\rho_c=\ell/g_c$. In both cases we take the same parameters
  $\alpha=1.3$, $\varepsilon_F=50\Omega$, and $\Lambda=10\Omega$. The
  lines are guides to the eye.}
\label{fig:tc}
\end{figure}

First, for frequencies close to the Debye energy,
$\omega\approx\Omega$, an additional feature of the density of states
starts to develop, and its strength grows with the magnitude of
$\lambda$. This is very similar to the phonon features in the density
of states of strong-coupling superconductors.  We emphasize, however,
that our theory predicts that the typical frequencies of phonons
coupled to electrons can be measured already in the normal,
non-superconducting state of a strongly disordered metal.

Second, when $\lambda$ increases, the dip in the density of states at
the Fermi level weakens. This effect is due to an anticorrelation
between the effects of the Coulomb pseudopotential and of the
electron-phonon coupling: increasing $\lambda$ diminishes the
self-energy $\chi(\varepsilon)$, see Fig.~\ref{fig:anticorrelation},
while increasing $\mu(\varepsilon)$ diminishes the self-energy
$\Sigma(\omega)$, see Fig.~\ref{fig:Z}. Looking at the Eliashberg
equations Eqs.~(\ref{eq:sigma},\ref{eq:chi}), the origin of the
anticorrelation can be traced back to the simultaneous presence of
both $\Sigma(\omega)$ and $\chi(\varepsilon)$ in the denominators of
the right-hand sides of both equations. Both
Figs.~\ref{fig:anticorrelation} and~\ref{fig:Z} show, however, that
the anticorrelation is relatively weak and to a first approximation it
can be neglected.

\section{Superconducting state}
Finally, we address the main subject of this paper, namely
superconductors with a sizeable AA anomaly in their normal state. As
explained in the previous Section, the requirement of experimental
observability of AA-type anomalies forces us to concentrate on the
strongly disordered regime with the Coulomb pseudopotential described
by Eq.~\eqref{eq:coulomb_strong}. Unless stated otherwise, in our
numerical calculations we assume the same set of parameters as in the
previous Section: $k_F\ell=1.8$, the fine-structure constant
$\alpha=1.3$, the Fermi energy $\varepsilon_F=50\Omega$, and the
cutoff $\Lambda=10\Omega$.  For the electron-phonon coupling we take
$\lambda=1.0$, and the length scale $L_s\geq\ell$ is taken as a free
parameter corresponding to the sample resistivity $\rho$ via
$L_s=g_c\rho$.

The Eliashberg equations
Eqs.~(\ref{eq:sigma},\ref{eq:chi},\ref{eq:phi},\ref{eq:psi}) have been
solved numerically. In a clean system with a constant Coulomb
pseudopotential $\mu(\varepsilon)=\mu_0$, our choice of parameters
leads to a reasonable critical temperature $T_{c0}\approx 0.033
\Omega$.  With increasing disorder, $T_c$ drops and, when entering the
strongly disordered regime, $T_c\approx 0.02 \Omega$. Further decrease
of $T_c$ as a function of $L_s/\ell$ in the strongly disordered regime
is shown in Fig.~\ref{fig:tc}. In the same figure we also plot $T_c$
for a somewhat smaller electron-phonon coupling constant,
$\lambda=0.8$.

An unexpected observation is that, although the insulating state is
approached as $L_s\rightarrow\infty$, the critical temperature does
not drop to zero in this limit and it stays constant.  Of course, the
mean-field Eliashberg equations can not be quantitatively correct for
$L_s\rightarrow\infty$, since fluctuation effects should be large
close to the insulating phase. Nevertheless, our fermionic theory is
certainly consistent with a direct superconductor-insulator transition
in 3D, without any intervening metallic phase.

\begin{figure}[t!]
\includegraphics[width = 7 cm]{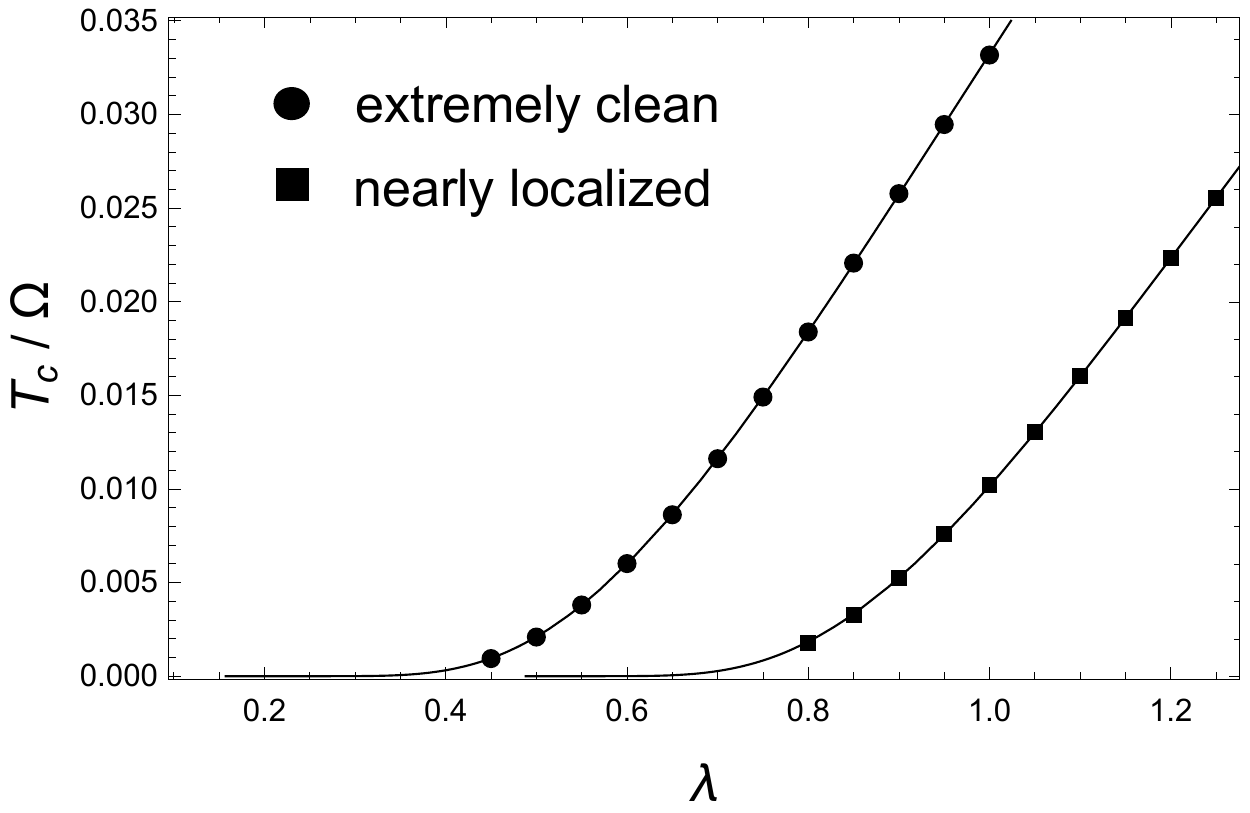}
\caption{Critical temperature $T_c$ as a function of the
  electron-phonon coupling $\lambda$ for extremely clean and nearly
  localized metals. The lines are fits described in the text.}
\label{fig:tc_lambda}
\end{figure}

\begin{figure}[b]
\includegraphics[width = 7 cm]{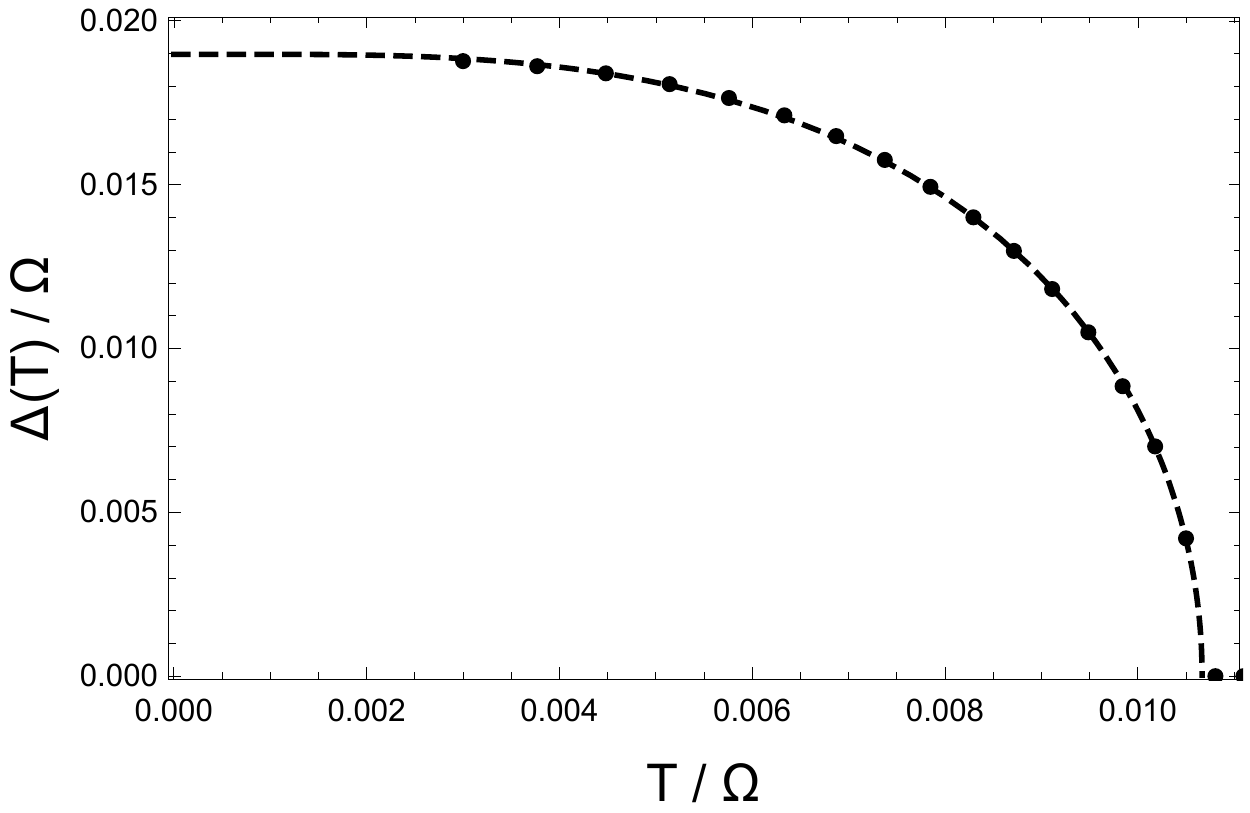}
\caption{Temperature dependence of the spectral gap
  $\Delta=\Phi(0,0)/Z(0)$ for an AA superconductor with
  $L_s/\ell=5$. The dots are numerical data and the dashed line is a
  fit described in the text.}
\label{fig:delta_T}
\end{figure}

In order to understand Fig.~\ref{fig:tc}, we have calculated the
critical temperature for a set of metals with fixed electronic
parameters and varying electron-phonon coupling $\lambda$. We have
considered two opposite limits for each $\lambda$: the metal was
assumed to be either extremely clean, $k_F\ell=10^8$, or nearly
localized, $k_F\ell=1.8$ and $L_s/\ell=7$.  The results are plotted in
Fig.~\ref{fig:tc_lambda}.  As expected, the critical temperature grows
with $\lambda$. Moreover, in both limits $T_c$ seems to be finite only
for $\lambda$ larger than a critical coupling strength
$\lambda_c$. This was also to be expected, since the phonon-mediated
attraction has to overcome the Coulomb repulsion.

Surprisingly, both data sets in Fig.~\ref{fig:tc_lambda} can be fitted
well by the simple formula $T_c=a\exp(-b/(\lambda-\lambda_c))$. From
these fits we estimate that in the extremely clean case the critical
coupling strength $\lambda_{c1}\approx 0.15$, whereas in the nearly
localized case $\lambda_{c2}\approx 0.46$. Note that
$\lambda_{c2}>\lambda_{c1}$, since the Coulomb repulsion is obviously
stronger in the nearly localized case.  It follows that three scenaria
for the metal-insulator transition are possible: (i) For
$\lambda<\lambda_{c1}$ the metal never becomes superconducting. (ii)
For $\lambda_{c1}<\lambda<\lambda_{c2}$ the metal can be
superconducting, provided it is sufficiently clean. Upon increasing
disorder, superconductivity disappears before entering the insulating
state \cite{note_KL}. (iii) For $\lambda>\lambda_{c2}$, all metallic
states become superconducting at low temperatures.

\begin{figure}[t]
\includegraphics[width = 7 cm]{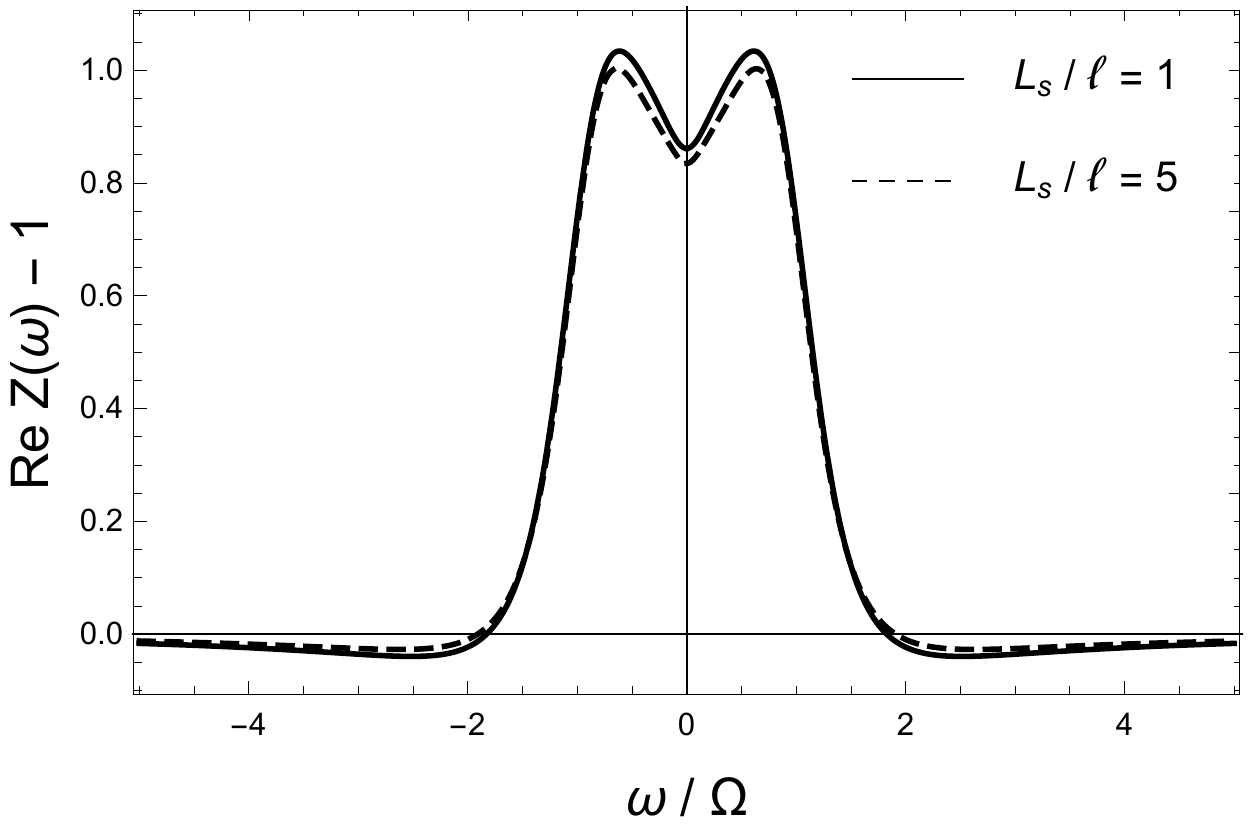}
\caption{The real part of the retarded wave-function renormalization
  $Z(\omega)$ for AA superconductors with $L_s/\ell=1$ and
  $L_s/\ell=5$ at temperature $T=0.003\Omega$.}
\label{fig:Z}
\end{figure}

\begin{figure}[b]
\includegraphics[width = 7 cm]{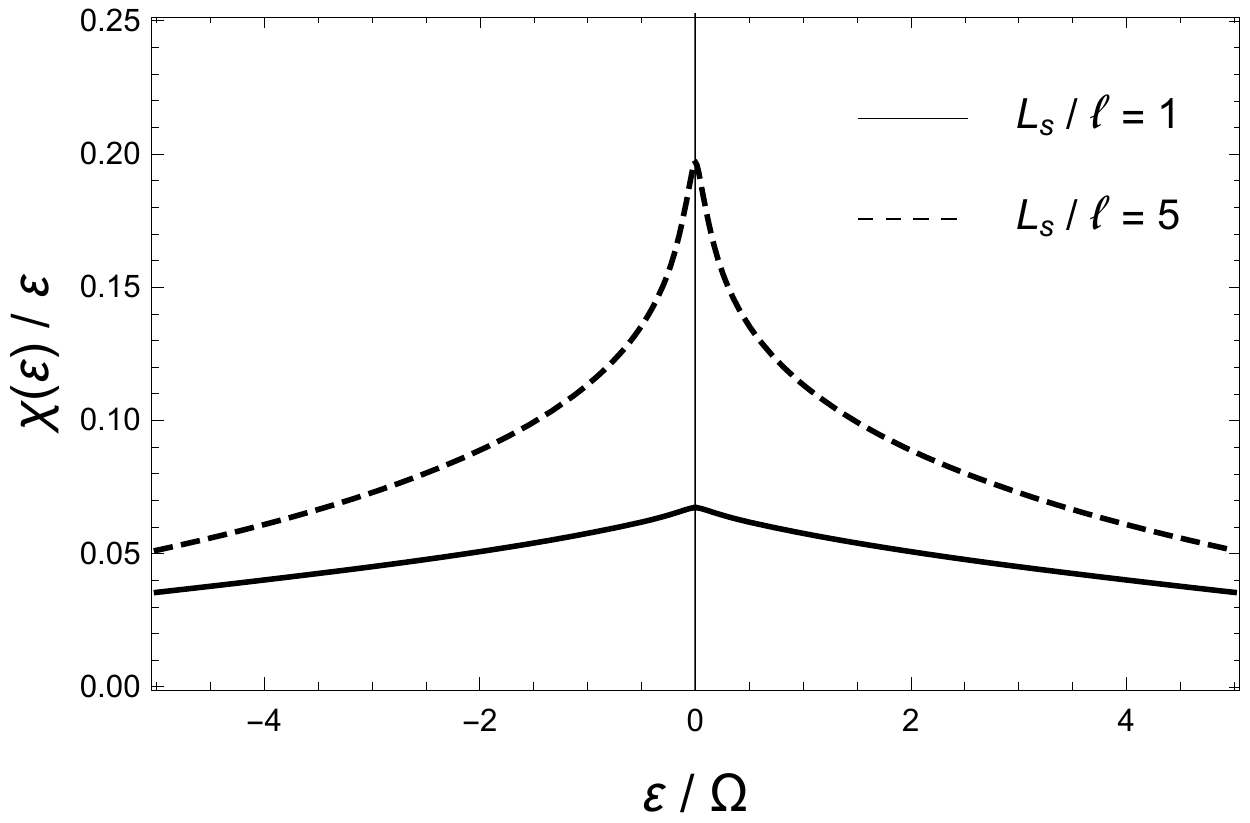}
\caption{Normalized self-energy $\chi(\varepsilon)/\varepsilon$ for AA
  superconductors with $L_s/\ell=1$ and $L_s/\ell=5$ at temperature
  $T=0.003\Omega$.}
\label{fig:chi}
\end{figure}

In the rest of this paper we concentrate on the physical properties of
strongly disordered AA superconductors.  In Fig.~\ref{fig:delta_T} we
plot the spectral gap $\Delta=\Phi(0,0)/Z(0)$ as a function of
temperature for an AA superconductor with $L_s/\ell=5$. We find that
the numerical data can be fitted well by the formula
$\Delta(T)=\Delta(0)\tanh[\alpha(T_c/T-1)^{1/2}]$, compatible with
simple BCS theory.  From the fit we obtain $\alpha\approx 1.77$,
$\Delta(0)\approx 0.0190 \Omega$, and $T_c\approx 0.0107 \Omega$,
implying that the ratio $2\Delta(0)/T_c\approx 3.55$, slightly smaller
than the clean-limit value for which we find 3.79.

In what follows we will compare the properties of two AA
superconductors, one with $L_s/\ell=1$, i.e. on the border between
weak and strong disorder, and another one with $L_s/\ell=5$,
i.e. deeply within the strongly disordered regime.

\begin{figure}[t]
\includegraphics[width = 7 cm]{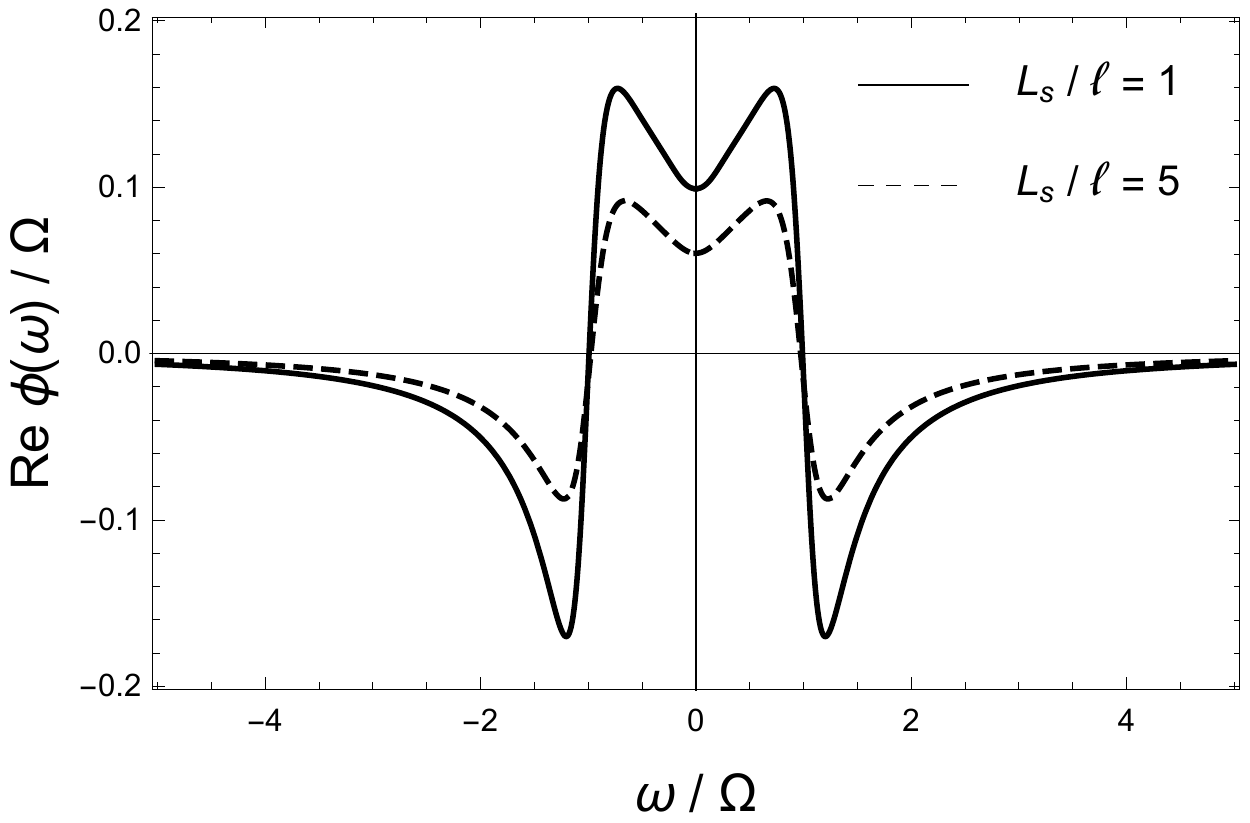}
\caption{The real part of the frequency-dependent anomalous
  self-energy $\phi(\omega)$ on the real frequency axis for AA
  superconductors with $L_s/\ell=1$ and $L_s/\ell=5$ at temperature
  $T=0.003\Omega$.}
\label{fig:phi}
\end{figure}

\begin{figure}[b]
\includegraphics[width = 7 cm]{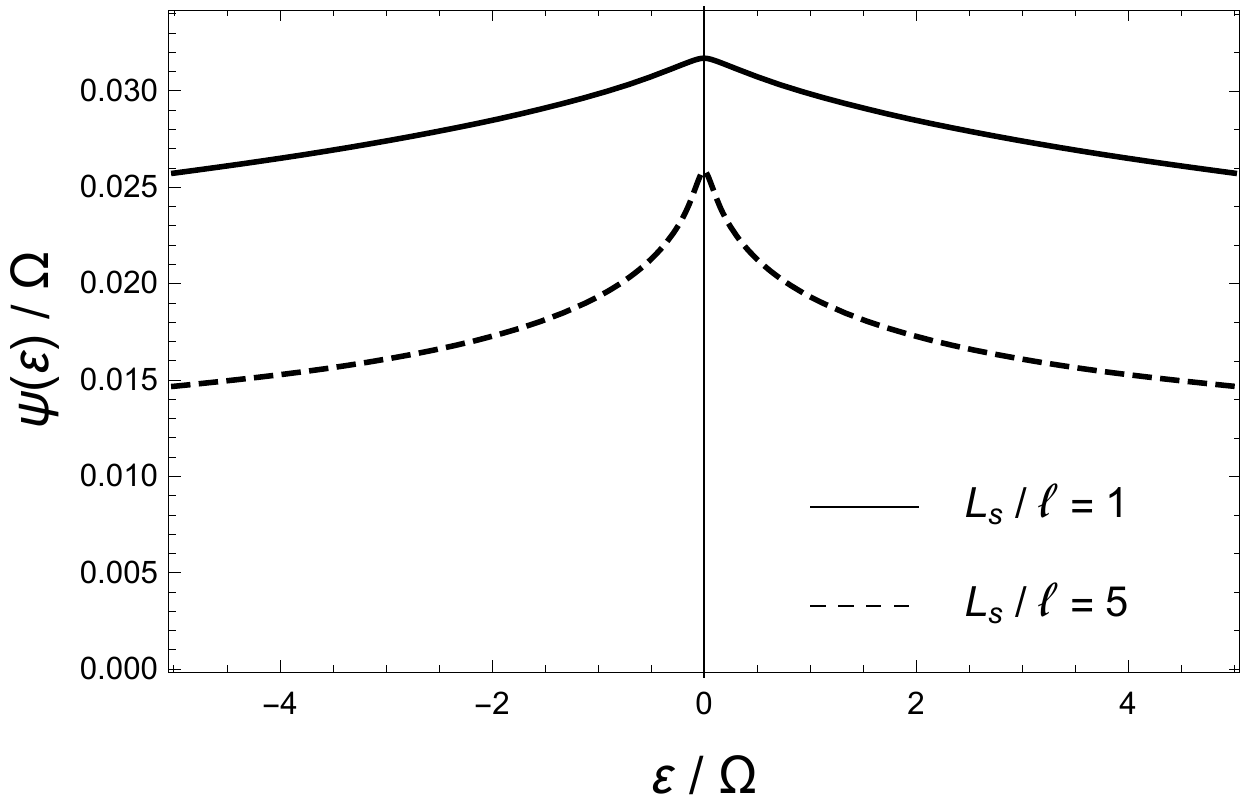}
\caption{The energy-dependent anomalous self-energy
  $\psi(\varepsilon)$ for AA superconductors with $L_s/\ell=1$ and
  $L_s/\ell=5$ at temperature $T=0.003\Omega$.}
\label{fig:psi}
\end{figure}

In Fig.~\ref{fig:Z} we plot the real part of the retarded
wave-function renormalization $Z(\omega)$, obtained by analytic
continuation from the imaginary axis. The overall shape of $Z(\omega)$
is in good agreement with the phonon-only analytic result.  One can
observe that the phonon-related function $Z(\omega)$ exhibits only
small changes with $L_s/\ell$. This is an example of the weak
anticorrelation between $Z(\omega)$ and the Coulomb pseudopotential
described in the previous Section.

\begin{figure}[t]
  \includegraphics[width = 7 cm]{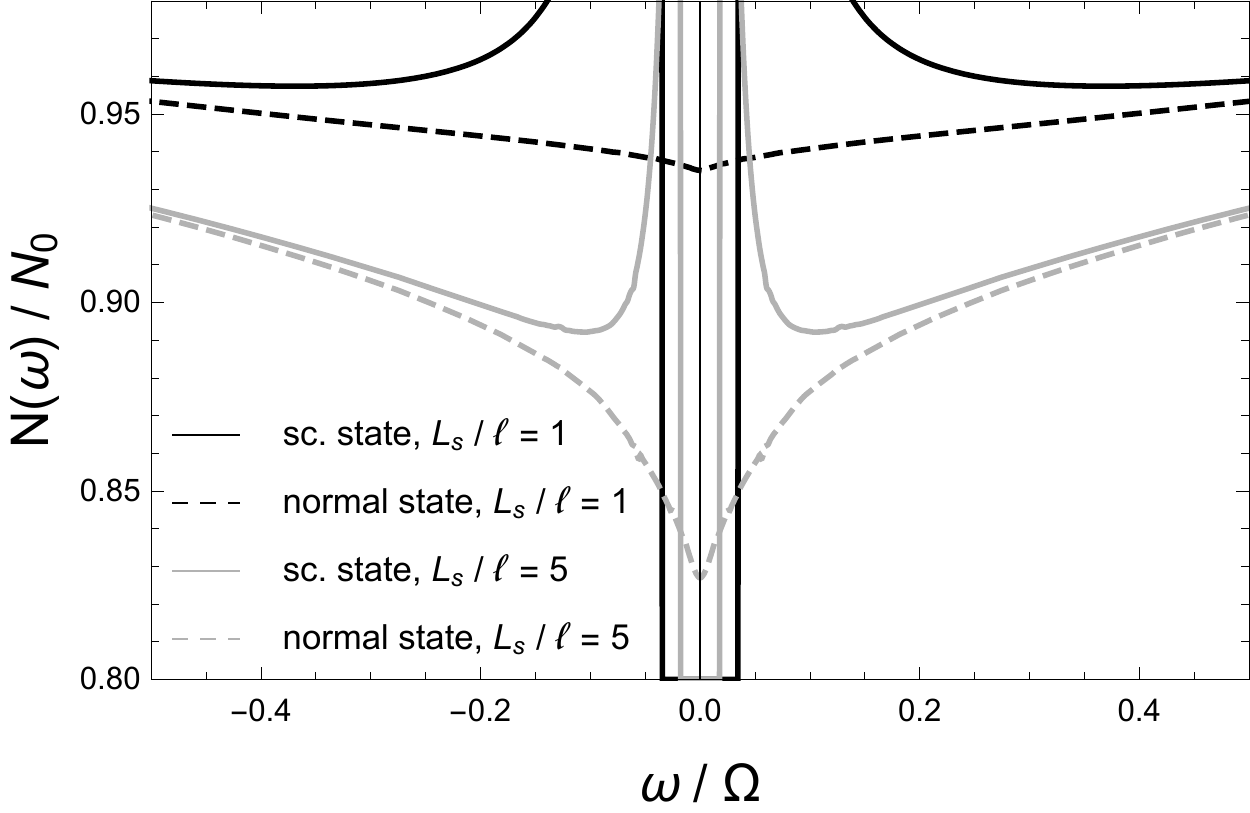}
\caption{Superconducting and (hypothetical) normal density of states
  for AA superconductors with $L_s/\ell=1$ and $L_s/\ell=5$ at
  temperature $T=0.003\Omega$.}
\label{fig:dos_sc}
\end{figure}

\begin{figure}[t]
  \includegraphics[width = 7 cm]{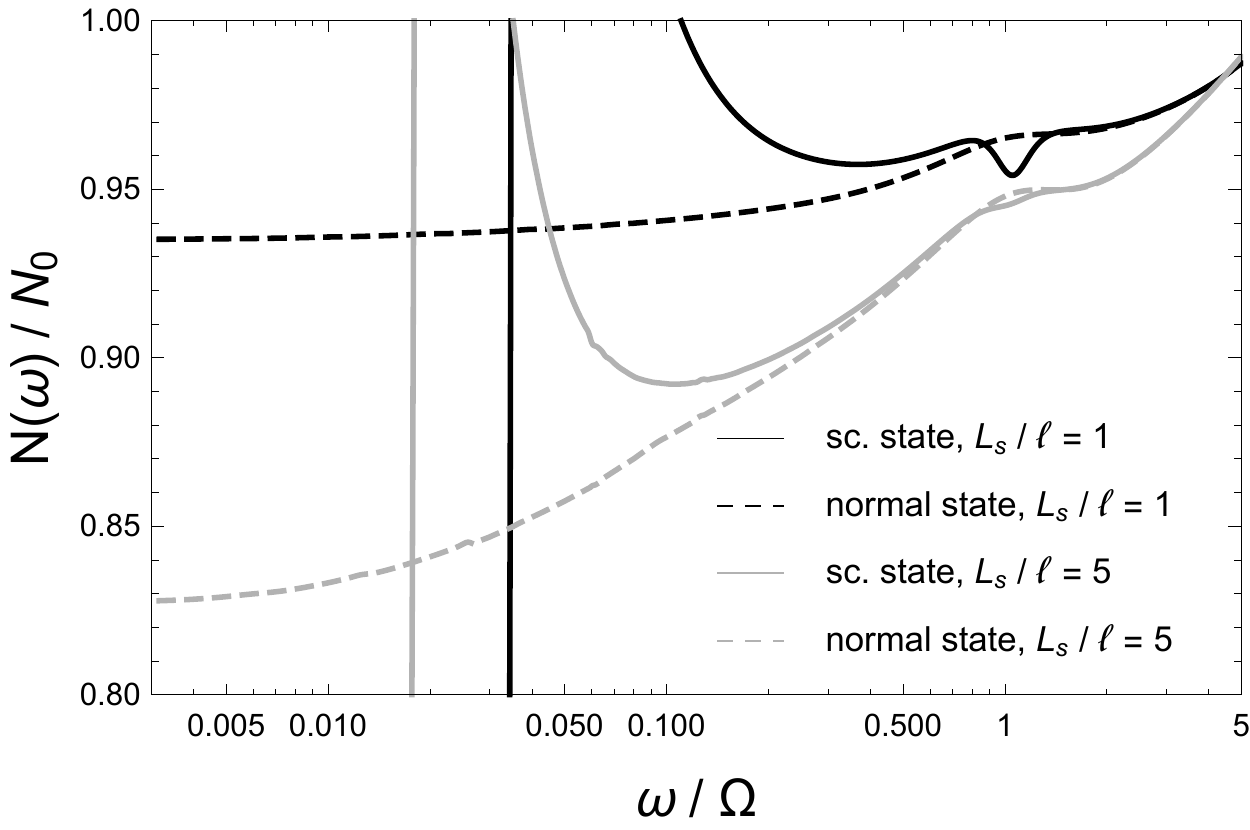}
\caption{The same as Fig.~\ref{fig:dos_sc}, but in a logarithmic scale
  of $\omega$.}
\label{fig:dos_log}
\end{figure}

On the other hand, as shown in Fig.~\ref{fig:chi}, the Coulomb
pseudopotential-related self-energy $\chi(\varepsilon)$ strongly
increases with increasing $L_s/\ell$. This was of course to be
expected.  As explained in Section~4, larger values of
$\chi(\varepsilon)$ imply stronger depression of the density of states
at the Fermi level in the hypothetical normal (non-superconducting)
state, see also Figs.~\ref{fig:dos_sc} and \ref{fig:dos_log}.

It is also worth pointing out that, as usual at moderate coupling, the
self-energies $Z(\omega)$ and $\chi(\varepsilon)$ change only little
between the normal and superconducting states.

The anomalous self-energy
$\Phi(\varepsilon,\omega)=\phi(\omega)-\psi(\varepsilon)$ is given by
the difference between an $\omega$-dependent part $\phi(\omega)$ due
to the phonons, and an $\varepsilon$-dependent part
$\psi(\varepsilon)$ due to the Coulomb pseudopotential. As shown in
Fig.~\ref{fig:phi}, the function $\phi(\omega)$ exhibits the standard
shape expected for boson exchange. When $L_s/\ell$ grows, the overall
scale of $\phi(\omega)$ decreases, but its shape remains roughly
intact.  This can again be interpreted as an anticorrelation effect.
In fact, the dominant effect of increasing $L_s/\ell$ is that
$R(\varepsilon)=1+\chi(\varepsilon)/\varepsilon$ grows, but from
Eq.~\eqref{eq:phi} it therefore follows that $\phi(\omega)$ has to
decrease.

The function $\psi(\varepsilon)$ is the superconducting analog of the
normal-state self-energy $\chi(\varepsilon)$. However, there is an
important difference between the two functions: for an
energy-independent Coulomb pseudopotential we have $\chi=0$, but
$\psi$ is a non-zero constant even in this case. If $\mu(\varepsilon)$
is not constant, then $\psi(\varepsilon)$ acquires a finite energy
dependence, as demonstrated in Fig.~\ref{fig:psi}. Note that with
increasing $L_s/\ell$, the energy depence of $\psi(\varepsilon)$
becomes more prominent.

A very rough estimate of the magnitude of $\psi$ can be obtained from
Bogoliubov's two-gap model: let us assume that $\phi(\omega)$ is a
finite constant for $\omega<\Omega$ and zero outside this interval,
and let $\psi(\varepsilon)$ be a constant up to the cutoff $\Lambda$.
Moreover, let us assume the presence of featureless electron-phonon
and Coulomb couplings $\lambda$ and $\mu$, where $\mu$ is an
appropriately taken average of $\mu(\varepsilon)$.  Then we find that
$\psi\sim (\mu^\ast/\lambda)\phi$, where
$\mu^\ast=\mu/[1+\mu\ln(\Lambda/\Omega)]$ is the renormalized Coulomb
pseudopotential. The data in Fig.~\ref{fig:psi} is roughly consistent
with this estimate.

Finally, the superconducting density of states of AA superconductors
with $L_s/\ell=1$ and $L_s/\ell=5$, calculated using
Eq.~\eqref{eq:dos_def}, is shown in Figs.~\ref{fig:dos_sc} and
\ref{fig:dos_log}. As expected, the pseudogap grows with increasing
$L_s/\ell$. Figure~\ref{fig:dos_log} shows that the phonon-related
peak at $\omega\approx \Omega$, visible already in the normal state,
acquires additional structure in the superconducting state, in
complete analogy with what is observed in superconductors with a
constant Coulomb pseudopotential.

\begin{figure}[b]
  \includegraphics[width = 7 cm]{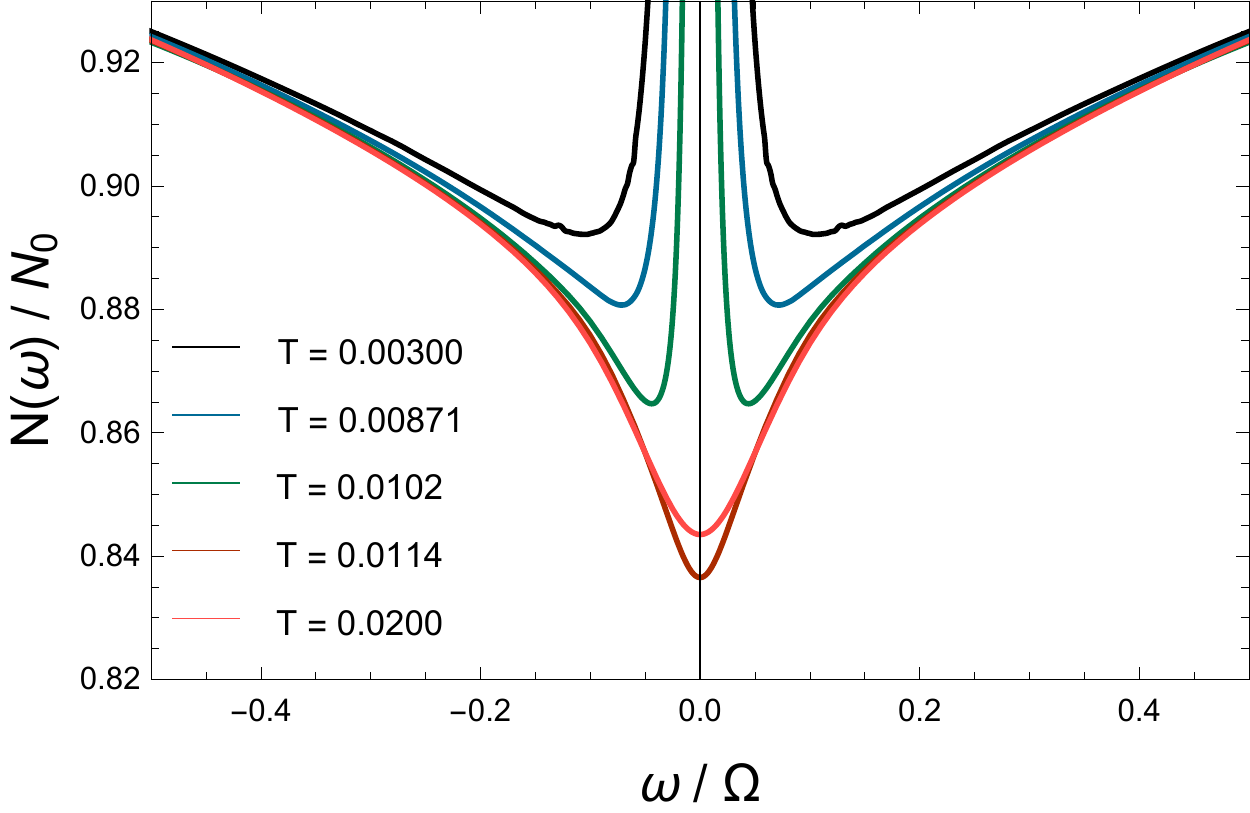}
\caption{Temperature dependence of the density of states of a strongly
  disordered AA superconductor with $L_s/\ell=5$. At temperatures
  below $T_c$, $N(\omega)$ is shown only for $|\omega|>\Delta$ for the
  sake of clarity.}
\label{fig:dos_T}
\end{figure}

In Fig.~\ref{fig:dos_T} we show the temperature dependence of the
density of states of a strongly disordered AA superconductor with
$L_s/\ell=5$. A pure $\sqrt{\omega}$ behavior takes place only for
$T\lesssim\omega\lesssim\varepsilon^\ast$, and at low energies the
normal-state singularity at $\omega=0$ is either cut off by the finite
temperature $T$ (above $T_c$), or completely masked by the
superconducting gap below $T_c$. Thus, weak-coupling superconductors
with $T_c\ll\Omega$ offer the most favourable conditions to
simultaneously observe both, the AA effect and the superconducting gap
in $N(\omega)$.

Let us comment on the relation of our theory to experiments. In this
work we have studied 3D superconductors and we have assumed that the
only effect of disorder is to introduce additional electron
scattering.  This means, however, that our theory can not be directly
compared with Refs.~\cite{Dynes81,Hertel83,Luna14,Luna15}.  In fact,
in order to interpret Refs.~\cite{Hertel83,Luna14,Luna15}, at the very
least it would be necessary to take into account the changes of the
electron density, but this is by no means straightforward, since not
only the fine-structure constant $\alpha$, but also $\varepsilon_F$,
$N_0$, and even $\lambda$ will change in that case. Similar
complications are to be expected also when varying the grain size of
granular aluminum \cite{Dynes81}. Moreover, large phase fluctuations
typical of the bosonic scenario are expected in the latter case.

It seems that the best example of AA superconductivity might be
provided by materials in which radiation damage causes large
resistivity enhancements, such as the $A15$ superconductors
\cite{Ghosh78,Putti08}. In order to check whether our picture for the
suppression of $T_c$ is valid in this group of materials, one should
start by looking for an AA anomaly in the normal state of the
high-resistivity samples. Tunneling data on the $A15$ compounds are in
fact available \cite{Kihlstrom84,Rudman84}, but unfortunately the
authors concentrate on the McMillan-Rowell inversion and do not report
the normal-state data \cite{note_inversion}. 

In order to estimate the order of magnitude of the changes in
$N(\omega)$ to be expected in the tunneling experiments, let us take
for the Debye frequency a typical value of $\Omega=$~40~meV, which for
our choice of parameters implies $\varepsilon_F=$~2~eV and
$T_{c0}\approx$~15~K.  From Fig.~\ref{fig:dos_T} it then follows that
in a strongly disordered AA superconductor with $L_s/\ell=5$ the
density of states at a temperature $T\approx T_c\approx$~5~K can
change by $r\approx$~12\% between $\omega=0$ and $\omega=$~20~meV, and
for the AA energy scale we get $\Delta_{\rm AA}\approx$~650~meV. These
estimates are quite similar to $r\approx$~14\% and $\Delta_{\rm
  AA}\approx$~520~meV, which have been measured for the $x=0.02$
sample of Ref.~\cite{Luna15}. Therefore such changes of $N(\omega)$
should be observable.

\section{Conclusions}
Building on the pioneering work of Belitz \cite{Belitz87}, in this
paper we have developed a formalism which can deal with both, the AA
effect and superconductivity, on equal footing. In particular, this
enables us to study the superconducting instability of systems with a
pseudogap caused by the AA effect in their normal state.

The set of generalized Eliashberg
equations~(\ref{eq:sigma},\ref{eq:chi},\ref{eq:phi},\ref{eq:psi}) has
been complemented by the simplest but physically well motivated
explicit expressions for the Coulomb pseudopotential in 3D,
Eqs.~(\ref{eq:coulomb_weak},\ref{eq:coulomb_strong}), and for the
phonon-mediated electron-electron interaction, Eq.~\eqref{eq:phonons}.
Following AMR \cite{Anderson83}, we distinguish between weakly and
strongly disordered conductors. In both cases, our expressions for the
Coulomb pseudopotential $\mu(\varepsilon)$ take into account the
$\sqrt{\varepsilon}$-type enhancement at the lowest energy transfers
$\varepsilon$.  In the strongly disordered case $\mu(\varepsilon)$ in
addition exhibits a logarithmic regime at intermediate energies,
predicted by AMR as a consequence of the anomalous
diffusion \cite{Anderson83}.

A complete numerical solution of the imaginary-time Eliashberg
equations has been presented, with emphasis on the parameter values
representative of strongly disordered AA superconductors.  This point
is crucially different from the approach of Belitz
\cite{Belitz87,Belitz89}, who postulates a simple functional form for
the self-energy $\chi(\varepsilon)$ which does not allow for the AA
anomaly, and solves the Eliashberg equations in the simple
two-square-well approximation.

Keeping the full energy dependence of the self-energy
$\chi(\varepsilon)$ allows us to show that the low-frequency behavior
of the density of states in the normal (non-superconducting) state is
well described by Eq.~\eqref{eq:dos_aa}. We have also calculated the
energy scale $\Delta_{\rm AA}$ and the density of states at the Fermi
level $N(0)$ in both, the weakly and strongly disordered regimes. In
agreement with earlier work \cite{Lee82,Muttalib83}, we find that the
AA anomalies are best observable in the strongly disordered
regime. When the electron-phonon coupling is turned on, we find (still
in the normal state!) additional anomalies in the density of states at
$\omega\approx\Omega$, where $\Omega$ is the Debye energy.

Numerical solution of the Eliashberg equations suggests there are two
possible scenaria for disorder-controlled superconductor-insulator
transition. If the electron-phonon coupling is weak, then the
transition proceeds via an intermediate metallic phase, in agreement
with the approximate theory of AMR \cite{Anderson83}.  On the other
hand, for sufficiently strong electron-phonon coupling, the transition
occurs without any intermediate phases.  It should be pointed out,
however, that the superconducting state in the vicinity of the
insulator is presumably fragile, and sufficiently strong {\it
  extrinsic} pair-breaking \cite{Herman18} may result in stabilization
of an intermediate metallic phase. If this happens, then the
transition again proceeds via an intermediate metallic phase.

A straightforward comparison of our results to experimental data is
not possible at the moment, since in all available experiments on 3D
disordered superconductors \cite{Dynes81,Hertel83,Luna14,Luna15}
introduction of disorder led, in addition to increased scattering,
also to a change of other relevant electronic parameters, such as the
Fermi velocity. In order to circumvent such difficulties, we have
instead proposed to search for the AA effect by tunneling spectroscopy
of radiation-damaged $A15$ superconductors. 

Our assumption that the electron-phonon coupling $\lambda$ does not
change with disorder is by no means obvious. The most complete
discussion of disorder-induced renormalization of the electron-phonon
coupling is due to Keck and Schmid \cite{Keck76}. These authors study
interaction between electrons and long-wavelength acoustic modes and
find that the coupling to longitudinal (transverse) modes decreases
(increases) with increasing disorder strength. Making use of these
results, Belitz argues that the total electron-phonon coupling
strength increases with disorder \cite{Belitz87}.  There are, however,
several caveats in this line of reasoning. First, when treating the
transverse phonons, Keck and Schmid consider only the so-called
collision-drag mechanism, and they neglect the electromagnetic
mechanism \cite{Fossheim02} with a different dependence on
disorder. Moreover, the effect of disorder on the phonons is not taken
into account. Second, to the best of our knowledge,
disorder-dependence of the coupling to the optical phonons has not
been studied yet. Since the electron-optical phonon interaction is
essentially due to electrostatics as for the longitudinal acoustic
modes, we expect a decrease of the coupling strength with increasing
disorder also in this case. Whether the total $\lambda$ increases or
decreases with disorder should therefore depend on the relative
contribution of the acoustic and optical phonons to the
electron-phonon coupling strength. Third, no systematic treatment of
the electron-phonon coupling taking into account the anomalous
diffusion of electrons is available at present.  Weak-localization
effects have been treated in the literature, but very different
conclusions have been reached: within a wave function-based approach,
it was argued that $\lambda$ decreases \cite{Park00}, whereas the
$\sigma$-model renormalization-group framework predicts an enhancement
of the interaction matrix elements \cite{Burmistrov12}.  Further work
is therefore clearly needed to arrive at definitive conclusions about
the disorder dependence of the electron-phonon coupling function
$g(\omega)$. In any case, the formalism developed in the present paper
will allow for a simple accomodation of the results of such studies in
a unified description of the Altshuler-Aronov effect and
superconductivity.

Further extensions of our theory are possible in several ways:  A
procedure analogous to the McMillan-Rowell inversion, but taking into
account the AA effect, should be worked out. 2D systems should be
studied, since in 2D one can make use of surface disorder which should
be free of the unwanted side-effects; moreover, high-quality data is
available in this case \cite{Sacepe10,Sacepe11}. Finally and most
ambitiously, it remains to be seen whether the AA effect proper or
some analogous effect play any role in the physics of the cuprates.

\begin{acknowledgments}
We thank P. Marko\v{s} and K. A. Muttalib for useful discussions.
This work was supported by the Slovak Research and Development Agency
under Contract No.~APVV-15-0496. B. R. was also supported by the
agency VEGA under Contract No.~2/0165/17. Numerical calculations were
performed at the Institute of Informatics of the Slovak Academy of
Sciences.
\end{acknowledgments}


\begin{thebibliography}{99}
\bibitem{note_rs}The more conventional parameter $r_s$ of the Coulomb
  gas is related to the fine-structure constant by
  $r_s=(9\pi/4)^{1/3}\alpha$.
\bibitem{note_ballistic}It is worth pointing out that in a system with
  a length-scale dependent ``diffusion'' constant $D(L)\sim v_F L$, the
  electron exhibits ballistic motion, since $L^2\sim D(L) t$ implies
  $L\sim v_Ft$.
\bibitem{note_KL}Both here and in (i), we neglect any possible
  Kohn-Luttinger instabilities, which are presumably very weak.
\bibitem{note_inversion} It is worth pointing out that if the AA
  effect turns out to be present, the standard inversion procedure
  does not apply, at the very least due to the fact that the
  superconducting state is described by two functions, $\phi(\omega)$
  and $\psi(\varepsilon)$.
\end{thebibliography}
\end{document}